\documentclass[12pt,a4paper]{article}
\usepackage[utf8]{inputenc}
\usepackage{amsmath}
\usepackage{amssymb}
\usepackage{graphicx}
\usepackage{booktabs}
\usepackage{longtable}
\usepackage{amssymb}
\usepackage{array}
\usepackage{subcaption}
\usepackage{hyperref}
\hypersetup{
    colorlinks=true, % false: boxed links; true: colored links
    linkcolor=black, % color of internal links (e.g., sections)
    citecolor=black, % color of citation links
    filecolor=black, % color of file links
    urlcolor=black   % color of external links
}
\usepackage{geometry}
\usepackage[
backend=biber,
style=chem-angew,
doi=false,
isbn=false,
url=false,
eprint=false
]{biblatex}

\newcolumntype{C}[1]{>{\centering\arraybackslash}p{#1}}

\title{Carbonic anhydrase II simulated with a universal neural network potential.}

\author{Tim Duignan, \\ 
\texttt{tim@orbitalmaterials.com} \\
Orbital Materials \\
School of Chemical Engineering, The University of Queensland}

\date{\today}

\begin{document}

\maketitle
\begin{abstract}
The carbonic anhydrase II enzyme (CA II) is one of the most significant enzymes in nature, reversibly converting CO$_2$ to bicarbonate at a remarkable rate. The precise mechanism it uses to achieve this rapid turnover remains unclear due to our inability to directly observe or simulate the full process dynamically. Here, we use a recently developed universal neural network potential (Orb) to simulate the active site of CA II. We reproduce several known features of the reaction mechanism, including the proton relay that conducts protons out of the active site to the His64 residue. Additionally, we observe a new reaction pathway where CO$_2$ reacts with a water molecule in the active site, which donates a proton to the zinc-bound hydroxide. This differs from the established mechanism where CO$_2$ directly reacts with hydroxide. Existing experimental data and independent quantum chemistry calculations are used to support the plausibility of this new mechanism. This demonstrates the potential of Orb to efficiently generate novel insights into important molecular scale processes that can potentially be harnessed to improve CO$_2$ capture technologies and drug design.  
\end{abstract}
%\keywords{Solutions, liquids,  ion diffusivities, thermodynamic properties, kinetic and structural properties, ab initio molecular dynamics, foundational neural network potentials.}

\section{Introduction}
Enzymes are nature's catalysts, orchestrating the chemistry of life with remarkable efficiency and specificity.  Their ability to accelerate reactions by many orders of magnitude under mild conditions has made them indispensable in both biological systems and industrial applications. However, our ability to harness and control these capabilities is limited by our inability to directly access the molecular scale processes that occur during these reaction events. 

While mutational and kinetic studies have provided valuable indirect evidence about enzyme function, they cannot directly reveal molecular-scale mechanistic details at atomic resolution. X-ray crystallography and cryo-electron microscopy offer high-resolution structural snapshots but only capture static conformations that may not represent the dynamic catalytic process. Density functional theory (DFT) calculations similarly provide static energetic insights but struggle with the complexity of large biomolecular systems.  NMR or IR spectroscopy can provide some dynamic information but typically only reports on specific labeled atoms and struggles with the size and complexity of most enzymes. This collective limitation in our experimental and computational toolbox has left significant gaps in our understanding of how enzymes achieve their remarkable catalytic efficiency.

This state of affairs is exemplified by the carbonic anhydrases (CAs) family of zinc-containing metalloenzymes. These catalyze the reversible hydration of carbon dioxide (CO$_2$) to bicarbonate (HCO$_3^-$) and a proton (H$^+$). CAs are ubiquitous across all kingdoms of life, and  play a vital role in various physiological processes, including respiration, pH regulation, bone resorption and ion transport. 

\begin{figure} 
      \begin{subfigure}[]{.47\textwidth}
        \includegraphics[width=\textwidth]{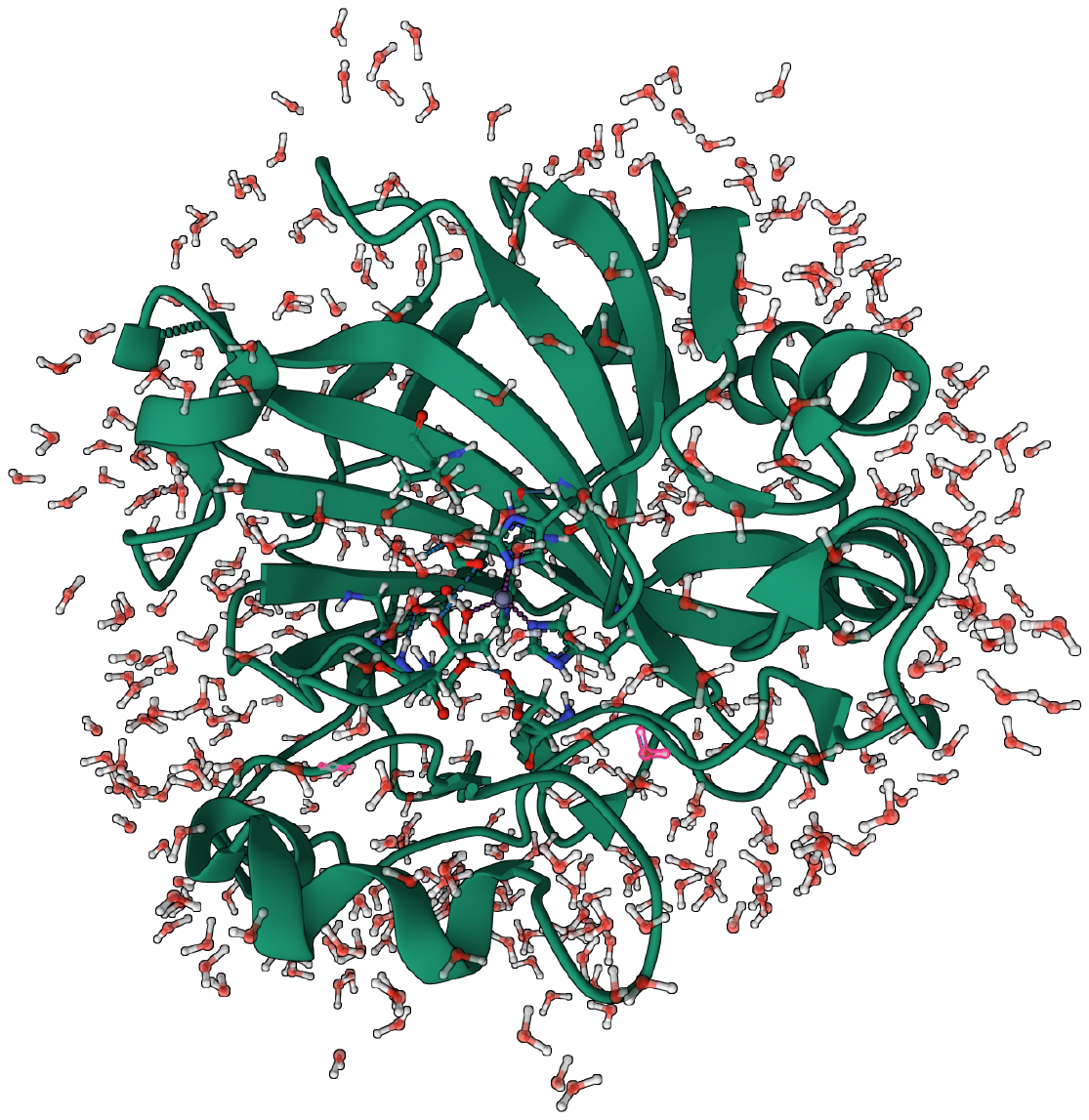}
    \end{subfigure}  
    \begin{subfigure}[]{.47\textwidth}
        \includegraphics[width=\textwidth]{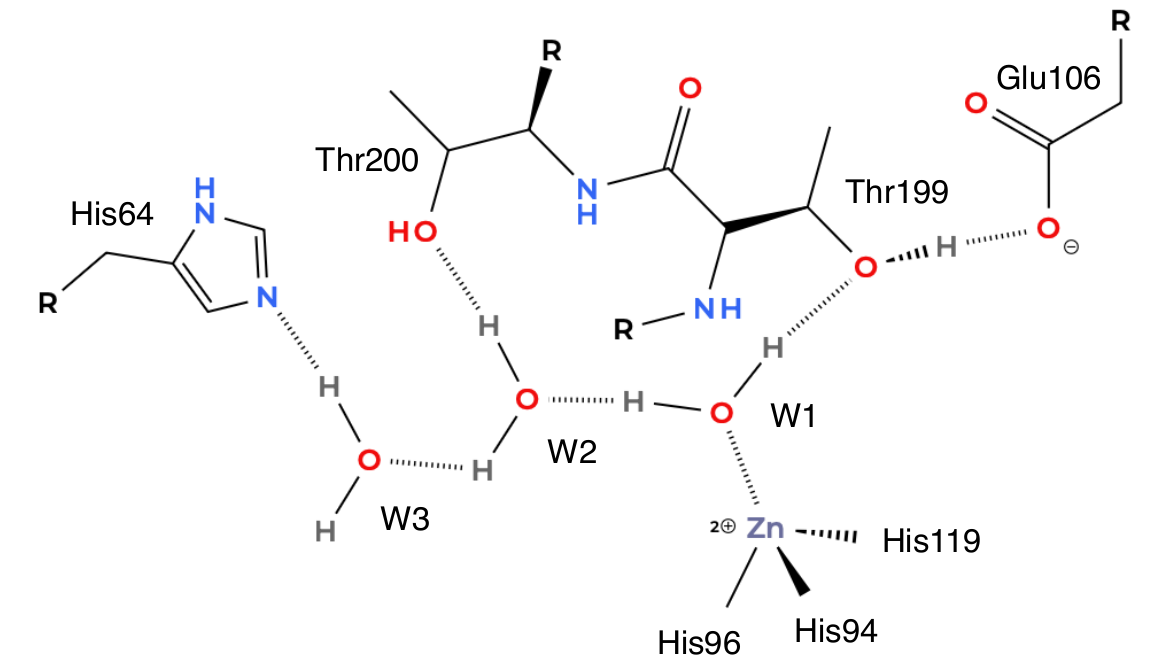}
        \end{subfigure}
        \caption{(a) A depiction of the Carbonic Anhydrase II enzyme. (b) A schematic of the active site demonstrating the key residues.  }
\label{fig:intro}
\end{figure}

CA II, (Figure~\ref{fig:intro}) specifically, achieves one of the fastest known catalytic rates in biology, with a $k_{\text{cat}}$ of $\sim10^6$ s$^{-1}$.\cite{jonssonPerspectivesClassicalEnzyme2020,kupriyanovaCarbonicAnhydraseUniversal2017} Due to this remarkable catalytic efficiency, CAs have garnered attention for potential applications in artificial lungs, biosensors, and CO$_2$ sequestration systems.\cite{booneCarbonicAnhydrasesTheir2013} In addition, drugs targeting CA II are used for treating glaucoma and certain types of cancer where pH regulation plays a crucial role.\cite{jonssonPerspectivesClassicalEnzyme2020}

Despite being one of the most extensively studied enzymes in biochemistry, with hundreds of crystal structures and decades of mechanistic investigations, our understanding of CA II's complete catalytic cycle remains fragmentary. The widely accepted mechanism involves a zinc-bound hydroxide nucleophilically attacking CO$_2$ to form bicarbonate, followed by bicarbonate displacement and regeneration of the zinc-hydroxide. 

However, crucial questions remain unanswered. For example, mutational studies have identified critical residues like Glu106, Thr199, and His64 that dramatically affect catalytic rates.\cite{desimoneExplorationResiduesModulating2019}   What mechanisms drive such loss in activity? Additionally, strong hydrogen kinetic isotope effects are observed for this reaction but the precise proton transfer network, essential for the enzyme's rapid turnover, has been particularly difficult to characterize dynamically. 

This knowledge gap is even more pronounced for thousands of other enzymes that have received far less scientific attention and resources. While CA II benefits from decades of focused research due to its medical and technological importance, the vast majority of the many thousands of human and microbial enzymes remain poorly characterized or completely unstudied at the mechanistic level. Many of these enzymes may harbor novel catalytic strategies that could inspire new technologies if we could observe their dynamic function in sufficient detail with a scalable automated strategy. 

Molecular dynamics simulations with standard biomolecular force-fields lack the ability to handle the chemical reactions critical to studying enzymes and so are of limited use for these applications.  While ab initio molecular dynamics or hybrid approaches such as QM/MM can describe reactive processes, these simulations typically handle only hundreds of atoms at quantum mechanical accuracy and are constrained to picoseconds timescales and hence cannot realistically access both the time and spatial scales necessary to model enzymatic function. 

The multistate empirical valence bond (MS-EVB) method  enables reactive MD simulations and has provided valuable insights into the CA II enzyme function. \cite{maupinElucidationProtonTransport2009,maupinProtonTransportCarbonic2010} However, this approach requires many parameters to be carefully tuned for specific moieties and the sensitivity to these specific choices remains unclear. 

Neural network potentials (NNPs), which use machine learning to approximate quantum chemical calculations, have emerged as a promising tool that bypasses the limitations of alternative approaches, offering quantum mechanical accuracy orders of magnitude faster than previously possible, approaching the speed of classical force fields.\cite{Behler2021,duignanPotentialNeuralNetwork2024} However, many current NNPs lack the ability to handle the diverse chemical space present in enzyme active sites, particularly in metalloproteins like carbonic anhydrase with its catalytic zinc center. Of the models that can handle the necessary chemical space most still struggle to access the relevant time and spatial scales. 

The recent development of Orb,\cite{neumannOrbFastScalable2024} a scalable and efficient universal neural network potential that combines a denoising pretraining objective with training on a large database of atomic forces computed with DFT on crystal structures, presents an opportunity to overcome these limitations and explore the dynamic behavior of metalloenzymes with unprecedented detail. We recently demonstrated the capability of Orb to provide insight into important biological phenomena by simulating the selectivity filter of the potassium ion channel, revealing previously unobserved new features. \cite{duignanPotassiumIonChannel2024} This capability is particularly surprising given the fact that Orb is trained purely on a database of small highly ordered crystal structures.\cite{jainCommentaryMaterialsProject2013,schmidtMachineLearningAssistedDeterminationGlobal2023a} However, the ability to provide novel insights into processes involving chemical reactions has yet to be demonstrated. 

Here, we present molecular dynamics simulations of carbonic anhydrase II using the Orb potential, producing nanosecond-scale simulations of systems containing thousands of atoms. Our simulations reproduce several known mechanisms by which this enzyme operates. In addition, we observed a new mechanism whereby CO$_2$ reacts with a water molecule in the enzyme pocket, which donates a proton to the zinc bound hydroxide. This differs from the established mechanism, where CO$_2$ reacts directly with the zinc bound hydroxide.  Evidence to support this mechanism from quantum chemical calculations and existing experimental evidence is provided.

These findings demonstrate the potential of Orb to advance our understanding of chemical reaction dynamics and inform the design of improved bio-inspired catalysts. The insights gained from these simulations could accelerate the development of new materials for CO$_2$ capture and conversion,\cite{appelFrontiersOpportunitiesChallenges2013,zhangCarbonicAnhydraseMembranes2022} while also providing valuable insights into diseases related to carbonic anhydrase. 

\section{Methods}
\subsection{System Preparation and Initial Structure}
\begin{figure}[tbh]
\centering
\includegraphics[width=1\columnwidth]{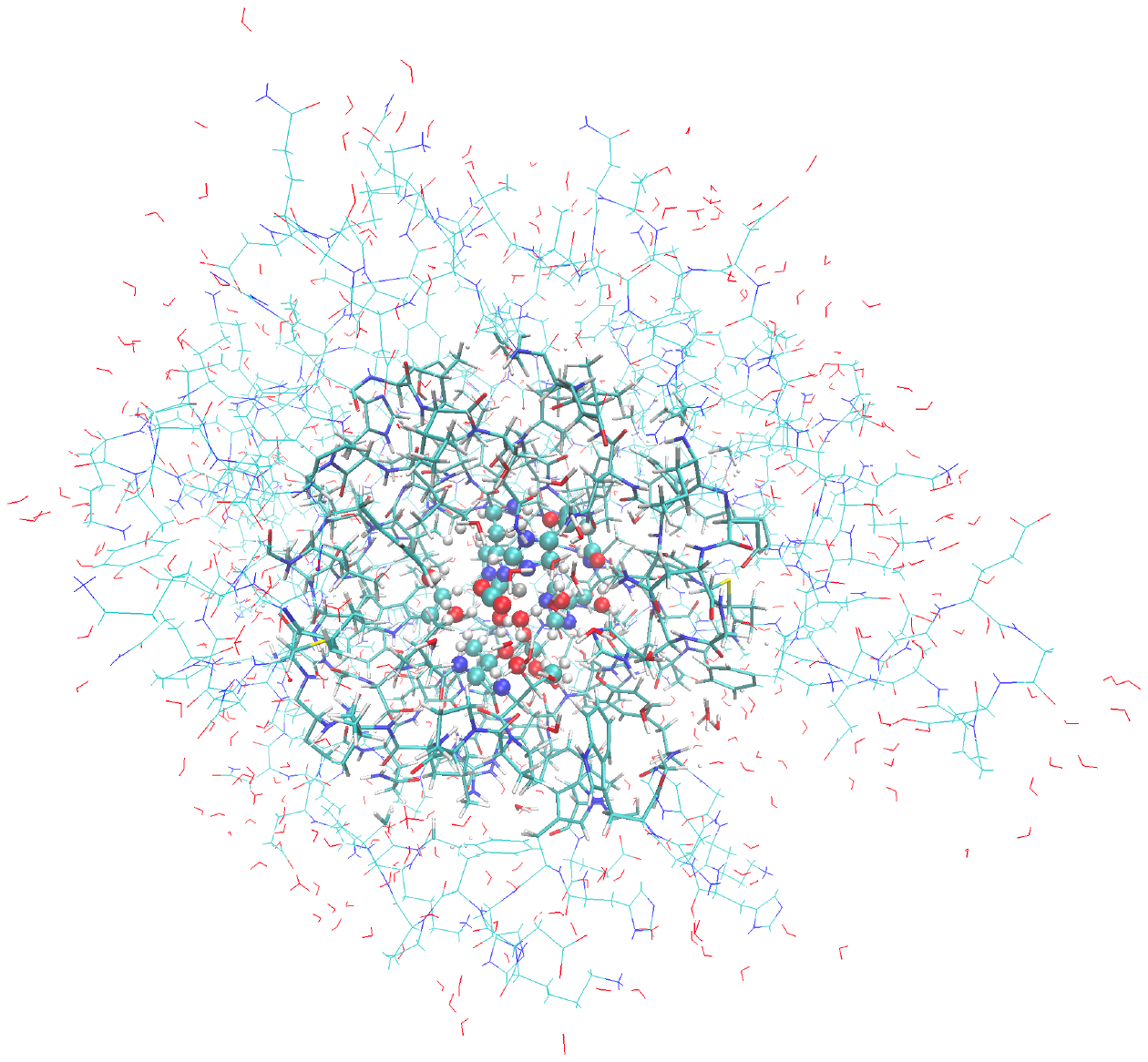}
\caption{A depiction of the Carbonic Anhydrase II enzyme simulation. The frozen outer region of the protein is represented with thin lines. The inner dynamic 15 \AA\ dynamic core with thicker lines and the active site itself with spheres. A  CO$_2$ and bound water molecules are included in the active site.}
\label{fig:fullview}
\end{figure}

The initial protein structure was obtained from the Protein Data Bank (PDB ID: 3ks3) and hydrogen's were added using pdb2pqr.\cite{liVeryFastEmpirical2005,sondergaardImprovedTreatmentLigands2011} The hydrogen on His119 had to be moved to the delta nitrogen to prevent clashing with the zinc. To reduce computational cost while maintaining essential dynamics only the PDB structure was used without additional solvation, i.e., only the water molecules present in the PDB structure were included. To keep the reactive pocket stable without solvating water molecules, atoms located more than 15 Å from the zinc center were frozen during the simulation. A CO$_2$ molecule was placed in an unoccupied position near the zinc center. To ensure sampling of potentially reactive configurations, a harmonic force (force constant of 5 eV/Å$^2$, cenetered at 5 \AA\ distance was applied between the Zn and CO$_2$.  This was only applied when the distance exceeded 5 \AA. Similar results were obtained with a force constant of  0.5 eV/Å$^2$ and 4 \AA, demonstrating the conclusions are not sensitive to these parameters.  A depiction of this simulation is provided in Figure~\ref{fig:fullview}.

\subsection{Molecular Dynamics Simulations}
Molecular dynamics simulations were performed using the Atomic Simulation Environment (ASE) package \cite{hjorthlarsenAtomicSimulationEnvironment2017} with the Orb universal neural network potential. The system was simulated using Langevin dynamics\cite{vanden-eijndenSecondorderIntegratorsLangevin2006} at physiological temperature (310 K) with a friction coefficient of 0.01. 

Simulations were run for 500 ps each. All reaction events were reproduced in at least three independent simulations.  

An automated reaction detection tool that identifies sustained changes in the atomic bonding graph and prints out trajectories of the corresponding time periods was used to automatically generate trajectories of the key chemical reactions observed in the simulation. A bond forming/breaking event is recorded if the actual distance between two atoms is greater/lower than half the sum of van der Waals radii for a sustained period, i.e., an average bonding graph is defined and when the difference to the fixed bonding graph exceeds 0.6 a new event is recorded. The bonding graph was updated every 100 fs and an exponential weighting average with $\alpha=0.1$ was used.

\subsection{Quantum chemistry}
Semi-empirical quantum chemical calculations were used to calculate the barrier height for the bicarbonate formation reaction with the GFN2-xTB \cite{bannwarthGFN2xTBAnAccurateBroadly2019} implemented with Rowan.\cite{RowanScientificHttps}  `Careful'  convergence settings were used. ($\Delta$ Energy: 0.000001 Ha;	Max Gradient: 	0.0009 Ha/\AA,	RMS Gradient: 0.0006 Ha/\AA). A model of the active site was created by including the three histidine residues that stabilize the zinc ion along with the Thr199, Thr200, Glu106, and His64 residues. The backbone atoms were restrained in place. 5 water molecules present in the active site were also included along with a CO$_2$ molecule. The 4 water molecules not included in the reaction had their oxygen atoms restrained. The structure was extracted from the Orb-MD simulation prior to the bicarbonate formation reaction. In total, 98 atoms were included.   The torsion drive algorithm was used for the constrained energy minimization.\cite{qiuDrivingTorsionScans2020}

The r2SCAN-3c\cite{Grimme2021} level of DFT was also used to recompute the energy barrier with a higher level of theory, using the structures obtained from GFN2-xTB.

\section{Results}
\begin{figure*}   \centering
      \begin{subfigure}[]{\textwidth}
        \centering
        \includegraphics[width=.95\textwidth]{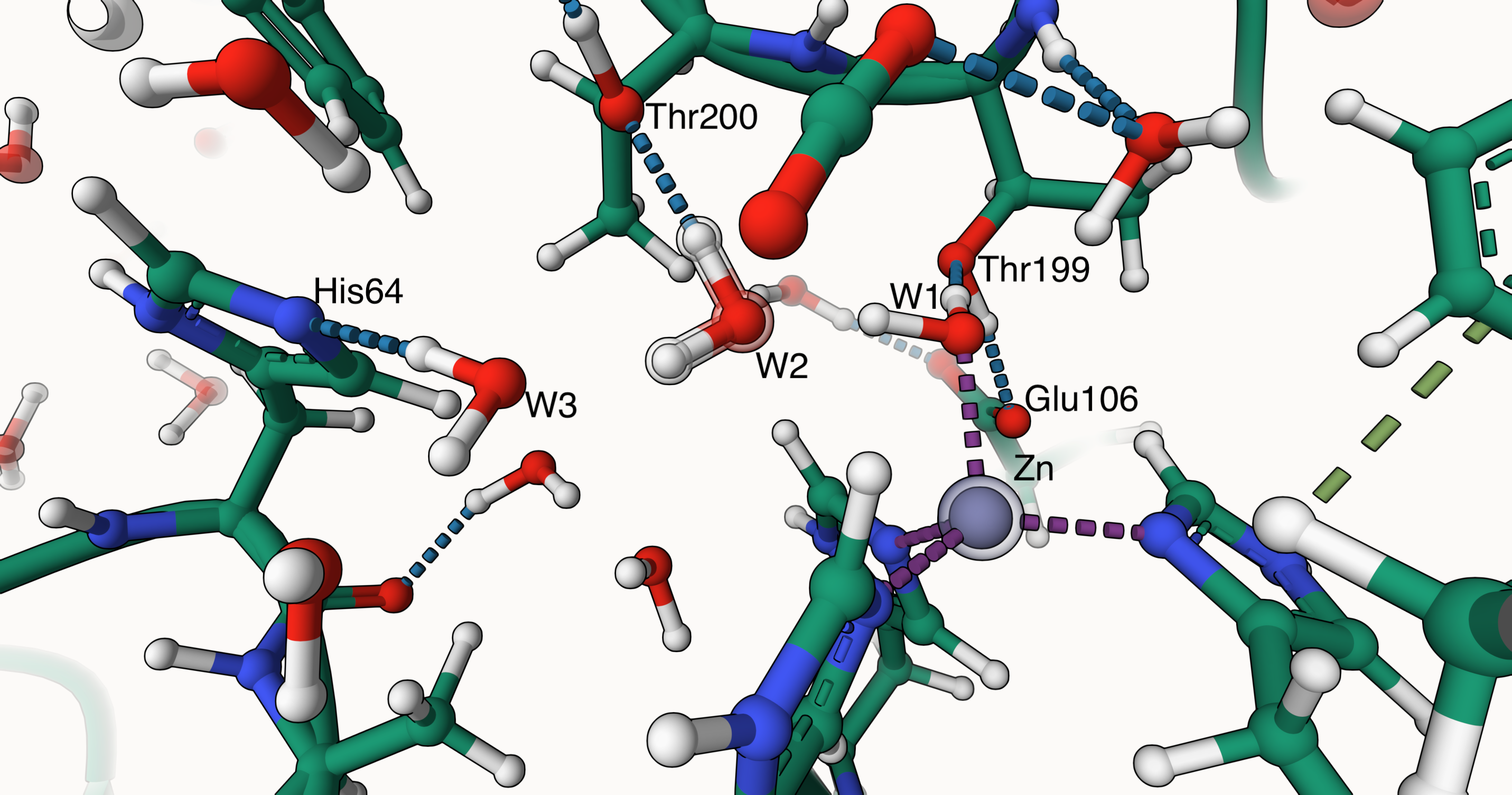}
        \caption{Initial structure}\label{fig:1}
    \end{subfigure}   \\
    \begin{subfigure}[]{\textwidth}
        \includegraphics[width=.95\textwidth]{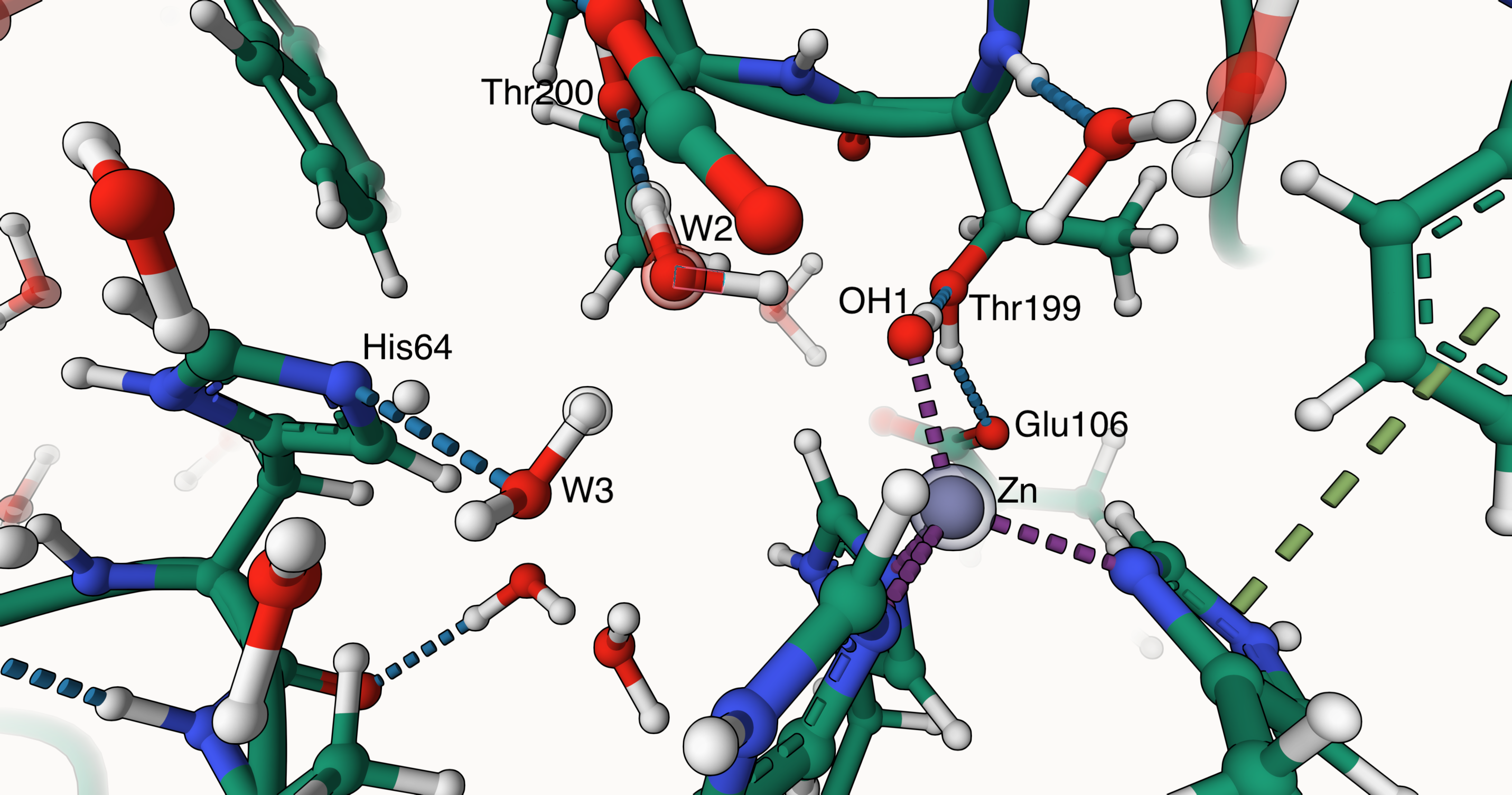}
        \caption{Hydroxide formation} \label{fig:2}\end{subfigure}\caption{(a) The initial structure of the active site with a CO$_2$ water molecule present. (b) Snapshot immediately after the formation of hydroxide by the three water proton relay. }
\end{figure*}

The simulations of the active site (Fig.~\ref{fig:1}) are very stable with no non-physical behavior observed, which is noteworthy given that Orb was trained exclusively on small inorganic crystal structures. This stability aligns with our previous observations of successful simulations of the potassium ion channel's selectivity filter.

Several simulations of this system were initialized. In three we observed a rapid proton hopping event that occurs spontaneously, where the water bound to the zinc converts to a hydroxide by donating a proton to the histidine 64 group through three rapid proton hopping events in a proton relay, using the Grotthuss mechanism see Fig.~\ref{fig:2}. This occurs in a semi-concerted mechanism, i.e., in two rapid steps, taking approximately 200 fs as previously proposed on the basis of static DFT simulations.\cite{cuiProtonWireConcerted2003}  Mutational studies also support this mechanism as they show that this His64 group is key to enabling the rapid turnover of CA II in the absence of a suitable buffer.\cite{tuRoleHistidine641989} Follow on proton hopping events were observed, where a single water molecule conducts the alpha proton from His64 to the His4 residue further from the active site.  

In an additional simulation, an alternative proton hopping event occurred where the proton on the zinc-bound water instead hopped onto the Thr199 residue, which donated its proton to Glu106. This proton relay has also been proposed previously.\cite{merzModeActionCarbonic1989} However, its role has remained unclear. The proton hops back and forth between the Glu106 and the zinc bound hydroxide over the course of the simulation. 

\begin{figure}[tbh]
\centering
\includegraphics[width=1\columnwidth]{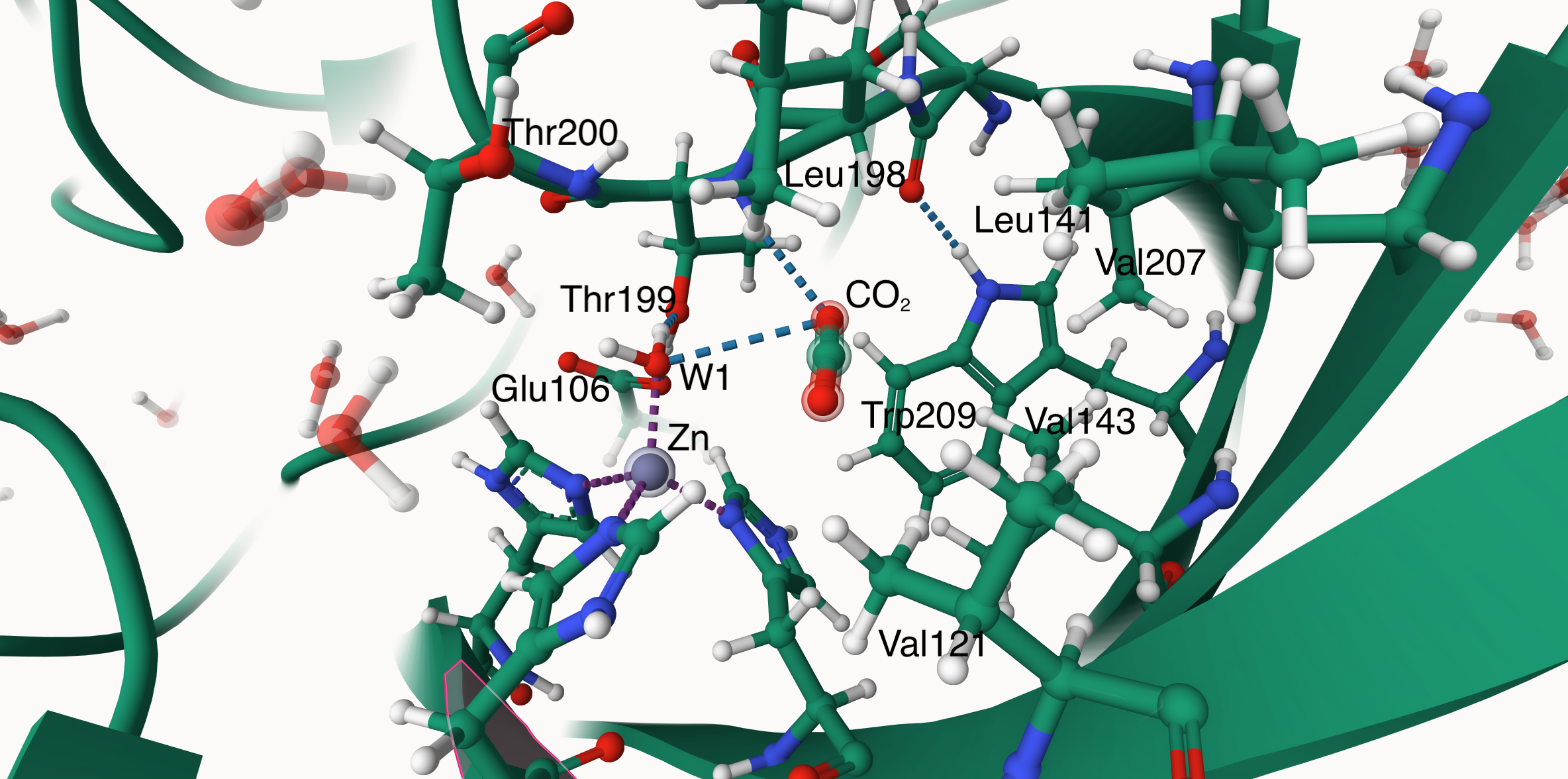}
\caption{A frame from MD simulations showing the CO$_2$ in its hydrophobic binding pocket in Carbonic Anhydrase II.}
\label{fig:CO2Pocket}
\end{figure}

Following this event the waters that were previously hydrogen bonded to the zinc-bound water migrate out of their positions. The zinc-bound hydroxide then accepts a hydrogen bond from Thr199 rather than donating one. This means it has a free H atom to donate a hydrogen bond. This appears to destabilise the surrounding water molecules which can no longer donate a hydrogen bond to the negatively charged hydroxide, as occurs after the proton has been conducted out of the active site. Instead, they must orient their oxygen towards the hydroxide to accept a hydrogen bond from it, meaning their dipole is oriented in an unfavorable direction. Possibly as a result, these water molecules were observed to diffuse away from the hydroxide. The CO$_2$ molecule in the active site was then observed to migrate into a hydrophobic binding pocket adjacent to the zinc bound hydroxide (Fig.~\ref{fig:CO2Pocket}) vacated by a water molecule. Crystal structures of CA II in high concentration of CO$_2$ have observed a CO$_2$ molecule sitting in this hydrophobic binding pocket.\cite{domsicEntrapmentCarbonDioxide2008,kimTrackingSolventProtein2016}   

Simulations of bicarbonate ion bound to the zinc were also run. Interestingly, the bond length between the hydroxide group and the carbon remained substantially larger than the other C-O bounds. (1.2 - 1.4 \AA\ vs. up to 1.7  \AA.). This has also been observed before in QM/MM simulations.\cite{lofererInfluenceBackboneConformations2003} Occasional formation of carbonate is observed by temporary transfer of a proton to Glu106. 

Applying a repulsive force between the carbon of the bicarbonate and the zinc of 0.1 eV/\AA\ yielded an interesting result: the C-OH bond quickly breaks, reforming CO$_2$ and OH$^-$. This implies a minimal barrier between CO$_2$ and the formation of bicarbonate. This has been previously observed in DFT studies. \cite{bottoniNewModelTheoretical2004}

\begin{figure*}   \centering
      \begin{subfigure}[]{.49\textwidth}
\includegraphics[width=\columnwidth]{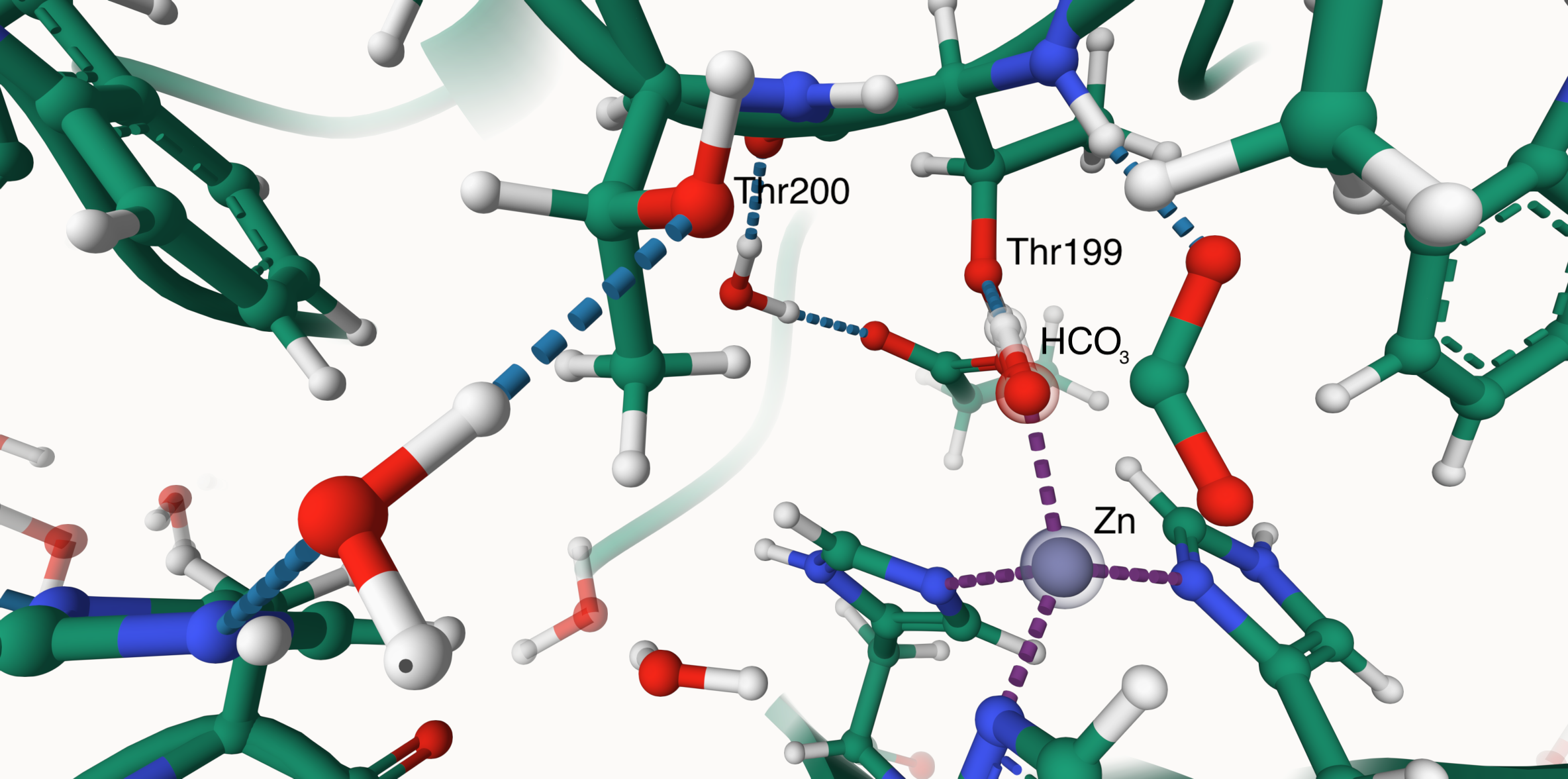}
\caption{Bicarbonate bound to zinc structure.}
\label{fig:HCO3}
    \end{subfigure}  
    \begin{subfigure}[]{.49\textwidth}
        \includegraphics[width=\textwidth]{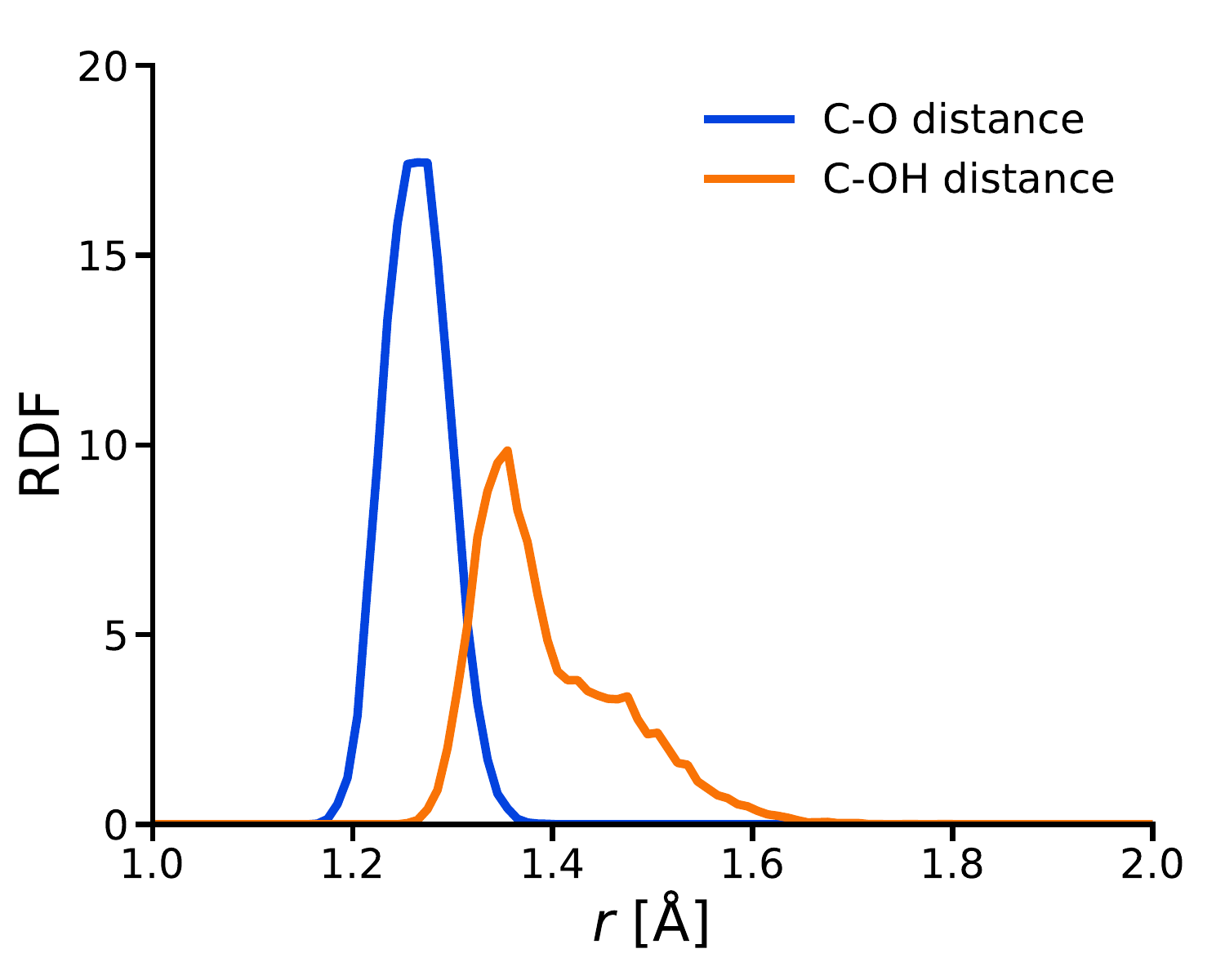}
        \caption{Carbon--oxygen RDFs.} \label{fig:gofrBicarb_comparison}
        \end{subfigure}
        \caption{(a) Bicarbonate bound to the zinc induced by adding a driving force. Notably, the C-OH bond is longer than the other two remaining bonds, indicating this is a hybrid between CO$_2$ and HCO$_3^-$. (b) RDF of the carbon of CO$_2$ and the zinc bound hydroxide oxygen vs. the CO$_2$ oxygen.}
\end{figure*}

In contrast, a force applied to the hydroxide oxygen merely rotated the HCO$_3^-$ without displacing it. These observations align with DFT-calculated barrier heights of 13 kcal/mol (or two barriers of 4 and 9 kcal/mol) for bicarbonate removal. This implies it may be easier for the bicarbonate to revert to CO$_2$ rather than to perform the next standard step in this mechanism which is the replacement of this HCO$_3^-$ with a water molecule.  There is no clear consensus on exactly how the bicarbonate is replaced by a water molecule due to a lack of experimental evidence. There are two published hypotheses for this mechanism, (Lipscomb vs. Lindskog)\cite{bottoniNewModelTheoretical2004}. We did not observe these hypothesized events in our simulations, even when applying a substantial force to the bicarbonate (both to the carbon or the oyxgen of the OH group). This suggests that an alternative pathway may be occurring, as this is known to not be the rate limiting step of the reaction. 

\begin{figure*}   \centering
      \begin{subfigure}[]{\textwidth}
        \centering
        \includegraphics[width=.94\textwidth]{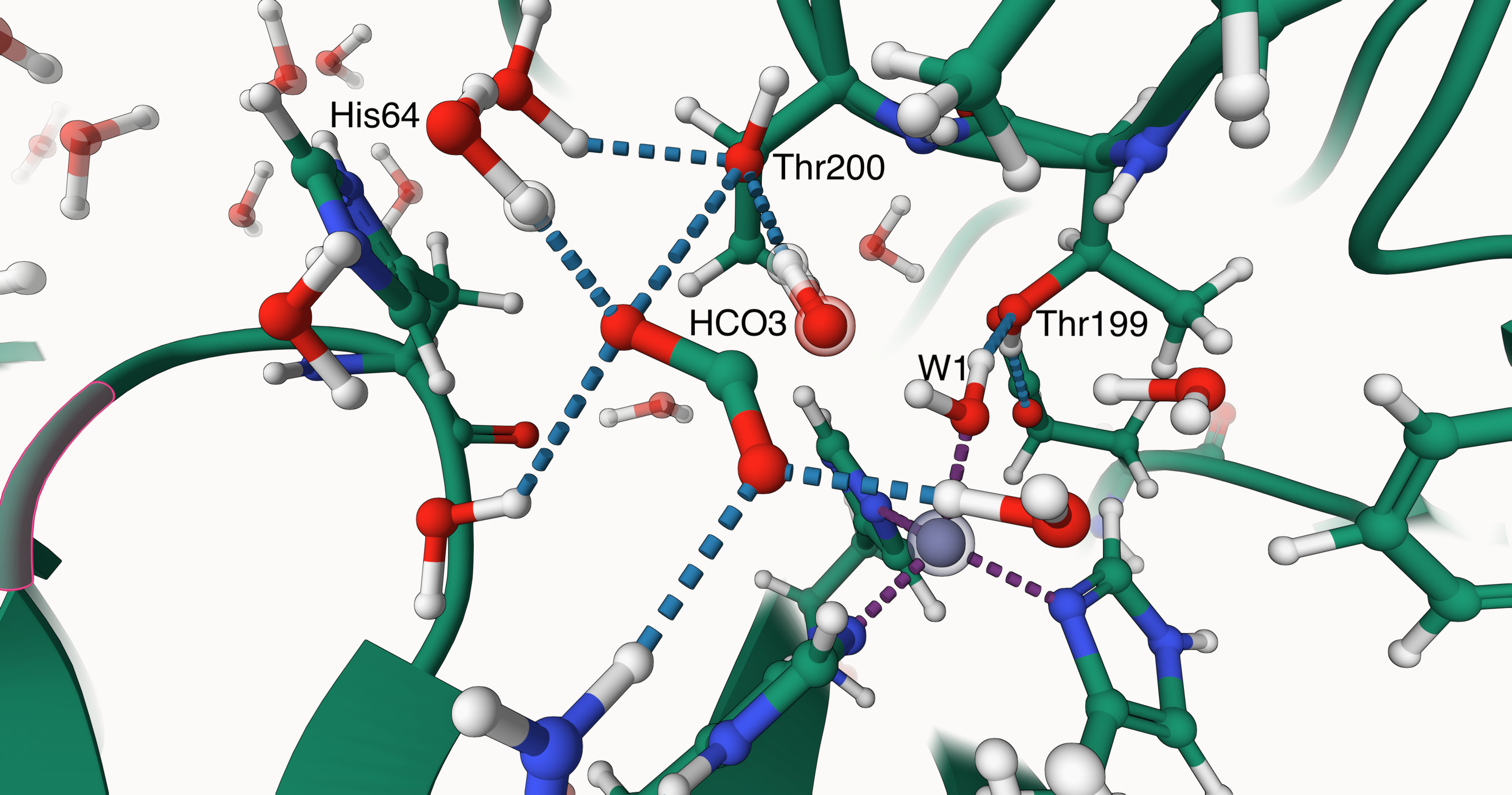}
        \caption{Bicarbonate formation}\label{fig:3}
    \end{subfigure}   \\
    \begin{subfigure}[]{\textwidth}
        \includegraphics[width=.94\textwidth]{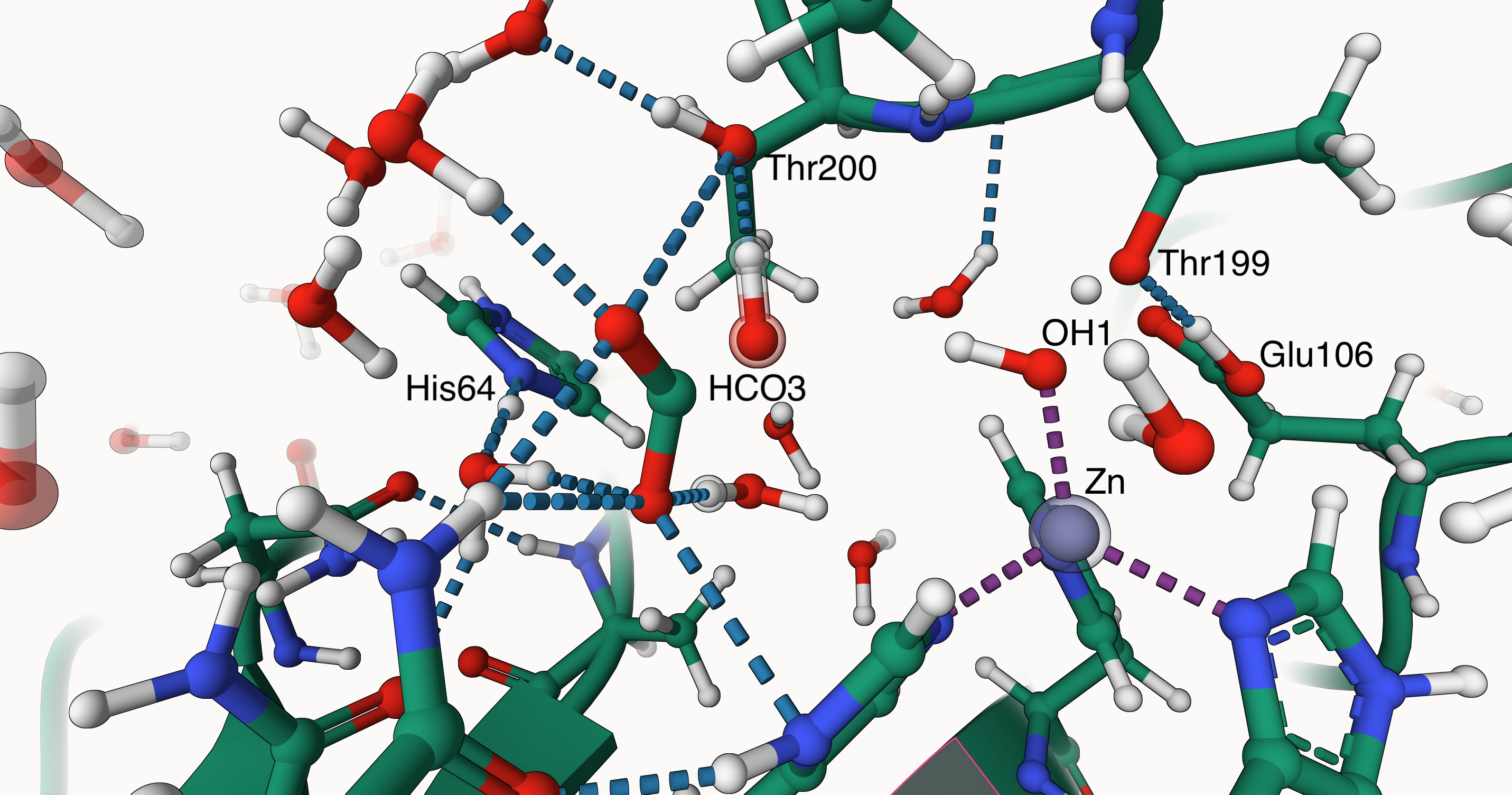}
        \caption{Hydroxide reformation.} \label{fig:4}
        \end{subfigure}
        \caption{(a) A Snapshot immediately after the formation of bicarbonate by reacting with the second water (W2), adjacent to the zinc hydroxide and THR200. (b) Soon after the formation of this structure, the hydroxide on the zinc is reformed by another proton relay, where the proton hops to Thr199 and then to Glu106.}
\end{figure*}

An alternative possible pathway emerges from running the original simulations for longer after the proton has hopped to His64, where the CO$_2$ was still constrained to be within 5 \AA\ of the zinc. In these simulations, CO$_2$ can occupy another position stabilized by some of the same hydrophobic residues that stabilise CO$_2$ in the known hydrophobic pocket.  This site is above the water (W2) bonded to the  zinc bound hydroxide  and is appearts to be more accessible from the surrounding environment. A reaction then occurs between this CO$_2$ and the water below it  (W2) to form bicarbonate (Figure~\ref{fig:3}). This is enabled by a concerted proton hop to the zinc bound hydroxide, i.e., the hydroxide acts as a base. The resulting water soon donates a proton to Thr199, which donates a proton to the Glu106 below as described above. (Figure~\ref{fig:4}) Once the constraint is removed from the carbon of the bicarbonate, it rapidly diffuses out of the binding pocket. 

\begin{figure*}  \centering
      \begin{subfigure}[]{.49\textwidth}
        \centering
        \includegraphics[width=\textwidth]{CAII-1.pdf}
        \caption{Step 1}\label{fig:CAII-1}
    \end{subfigure}   
    \begin{subfigure}[]{.49\textwidth}
    \centering
\includegraphics[width=\textwidth]{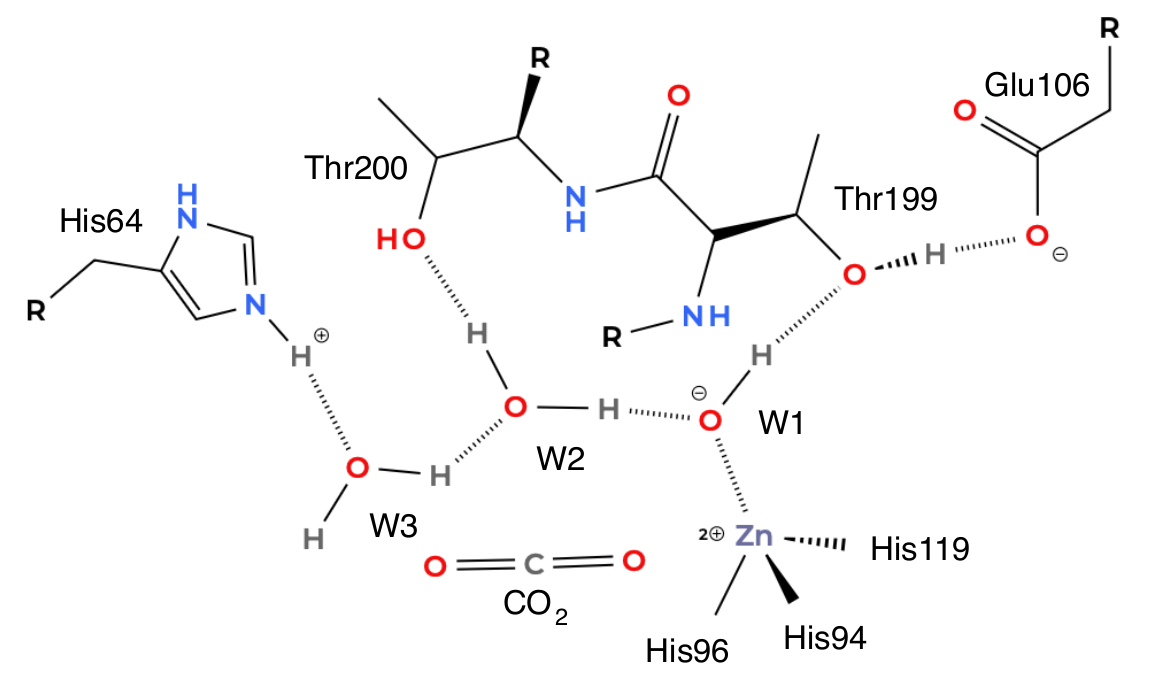}
        \caption{Step 2} \label{fig:CAII-2}
        \end{subfigure}\\
            \begin{subfigure}[]{.49\textwidth}
    \centering
\includegraphics[width=\textwidth]{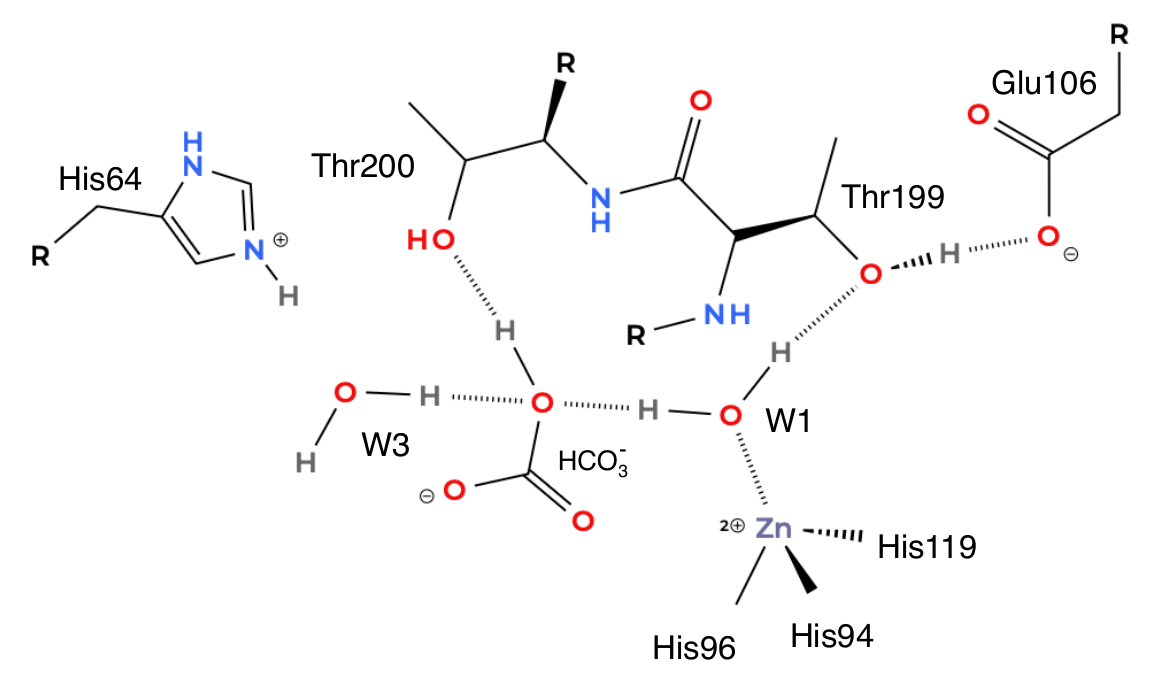}
        \caption{Step 3} \label{fig:CAII-3}
        \end{subfigure}
            \begin{subfigure}[]{.49\textwidth}
    \centering
\includegraphics[width=\textwidth]{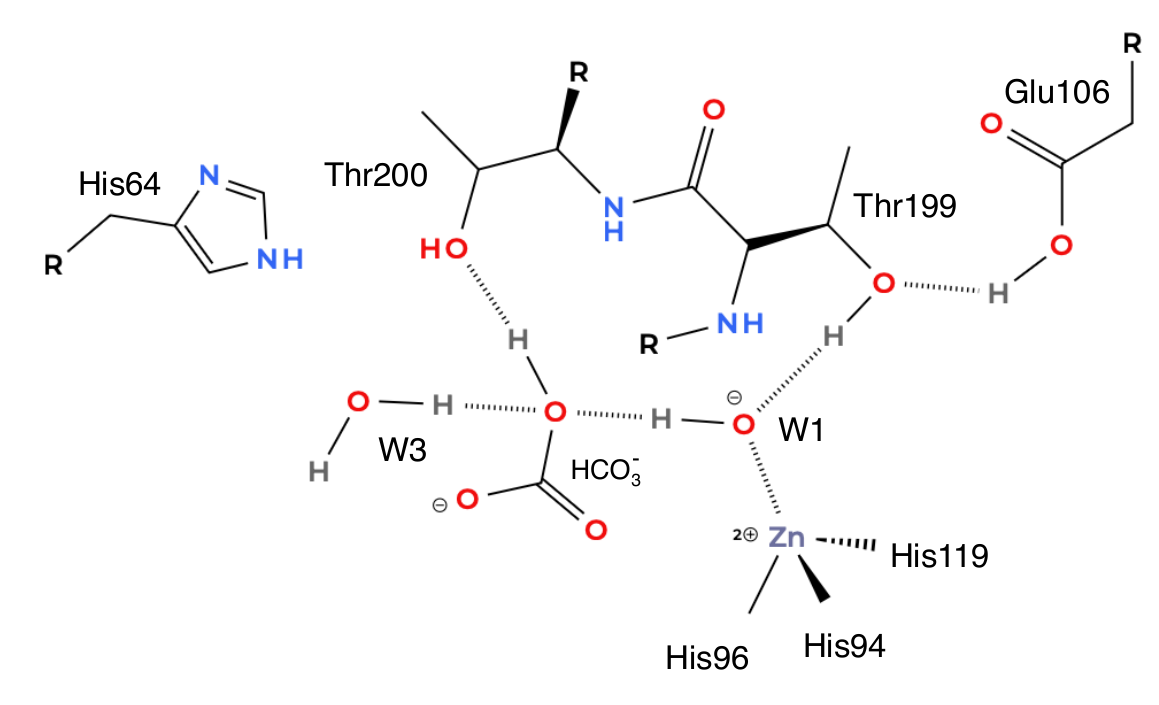}
        \caption{Step 4} \label{fig:CAII-4}
        \end{subfigure}
        \caption{Depiction of the four steps of the reaction mechanism observed in the simulations.}
        \label{fig:CAII} 
\end{figure*}

To the best of our knowledge such a reaction mechanism has not been proposed before and a general consensus that the hydroxide directly reacts with CO$_2$ has stood for many decades despite extensive research. The full reaction is depicted in Figure~\ref{fig:CAII}. 

\subsection{Quantum chemistry}
\begin{figure*} 
\centering
      \begin{subfigure}[]{.42\textwidth}\centering
\includegraphics[width=\textwidth]{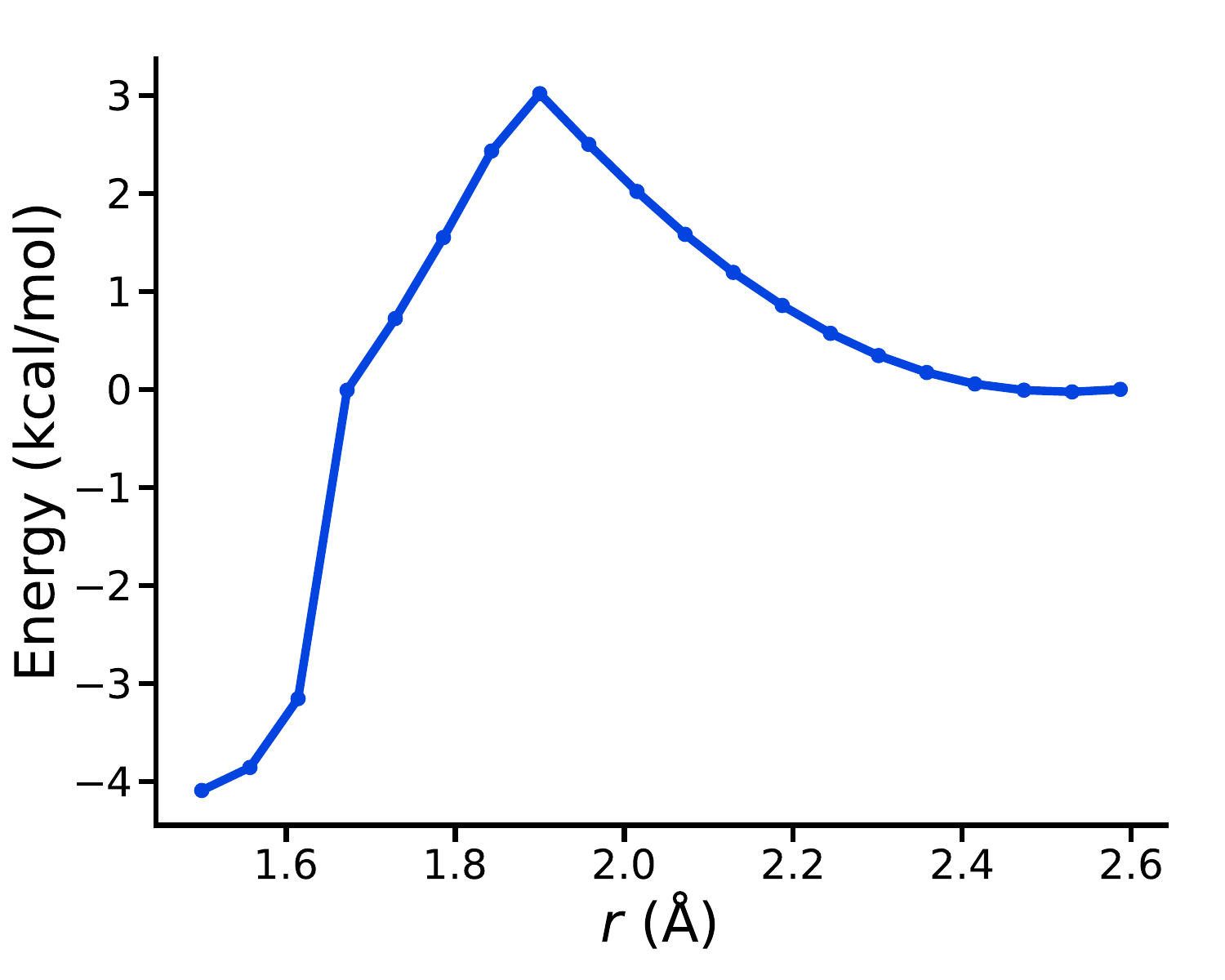}  
    \end{subfigure}   
   \begin{subfigure}[]{.42\textwidth}\centering
     \includegraphics[width=\textwidth]{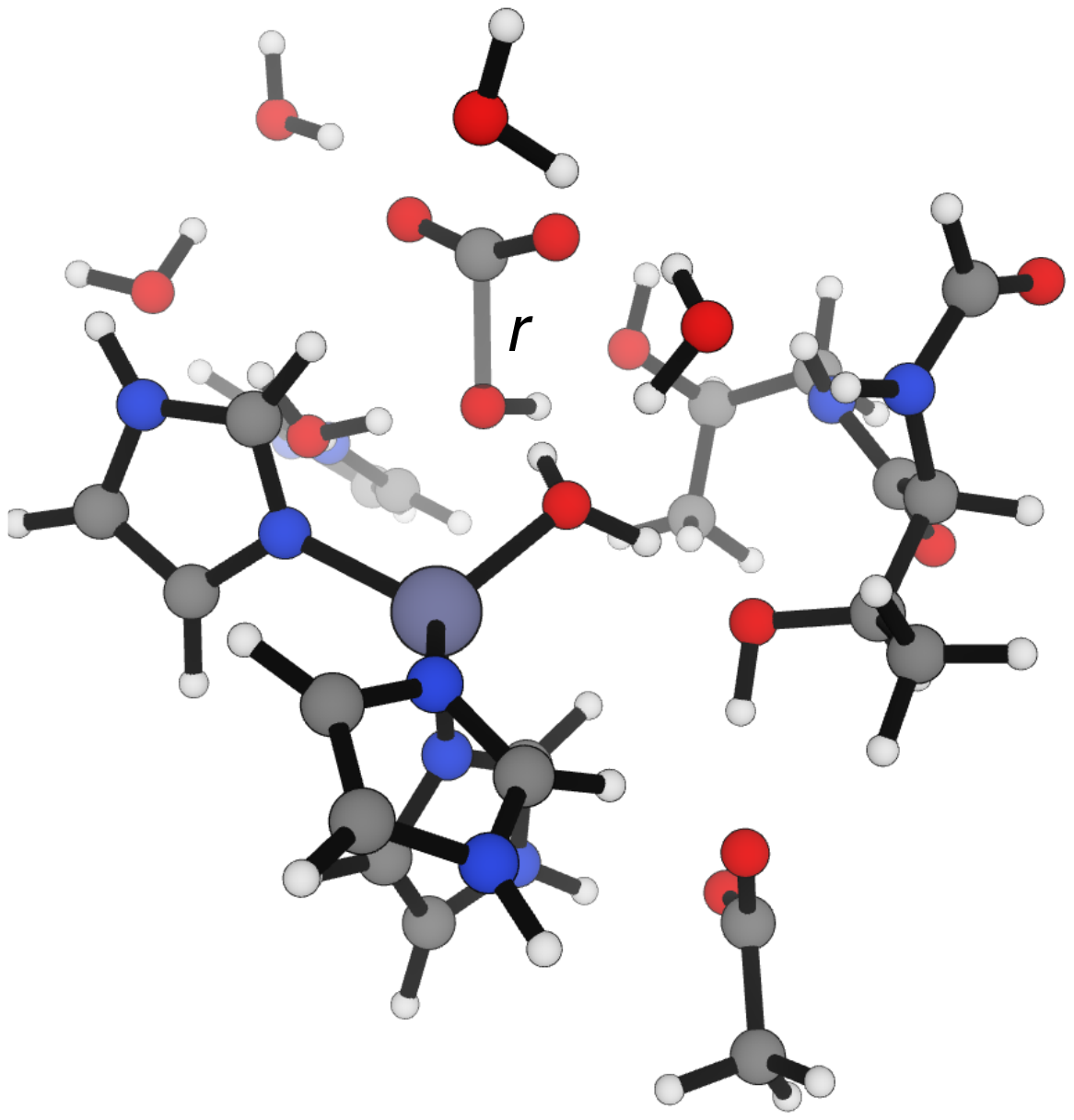}
        \end{subfigure}
        \caption{(a) Energy as a function of separation between the carbon of CO$_2$ and the oxygen of the water molecule (W2) in the active site. (b) Depiction of the active site used for the semi-empirical calculations, after the proton transport to the zinc bound hydroxide has occurred.} \label{fig:energy_barrier_plot}
\end{figure*}

As this system is far outside the training data distribution for Orb, it is important to independently confirm the plausibility of this novel reaction mechanism. Therefore we use a semi-empirical level of theory (GFN2-xTB) to compute the energy barrier for this process. The results are shown in Figure~\ref{fig:energy_barrier_plot}. A reaction barrier of 3 kcalmol$^{-1}$ is produced. Crucially, the hopping of the proton from the water molecules to the zinc bound hydroxide spontaneously occurs to prevent the energy barrier from raising any higher. The structure immediately after the proton hop occurs is also shown with a carbon oxygen separation of 1.843 \AA. 

Recomputing the energy barrier with the higher r2SCAN-3c\cite{Grimme2021} level of theory results in a barrier height of $8.3$ kcalmol$^{-1}$, which is higher but actually more consistent with the experimentally observed reaction rate of 10$^6$ s$^{-1}$. The fact that this reaction is observed more quickly than this in the Orb simulations is attributable to the significant constraining force applied to keep the CO$_2$ near the zinc center. 

\section{Discussion}
Orb's capability to reproduces several well-established features of the carbonic anhydrase II (CA II) reaction mechanism with almost no human guidance and moderate computational resources is a remarkable demonstration of its capabilities, and also justifies serious consideration of the newly observed reaction mechanism, where CO$_2$ reacts with a water molecule adjacent to the zinc-hydroxide complex rather than directly with the hydroxide itself.

This novel reaction pathway may be faster than the established one as it involves far fewer steps, with comparable barrier heights, i.e., high-quality  DFT studies examining the bicarbonate-to-water replacement at the zinc site reveal six steps with multiple transition barriers of 0.6, 4.4, 9, 3.8, and 8.6 kcalmol$^{-1}$.\cite{bottoniNewModelTheoretical2004} Additionally, this reaction pathway would be further slowed by the steps associated with the CO$_2$ diffusing into the binding pocket, which is buried $\approx$15 \AA\ deep within the enzyme and displacing the water molecules that are normally bound to the zinc-bound hydroxide. 

The new mechanism observed would also explain the why elaborate and carefully conserved nature of the active site is necessary, as it requires a precise arrangement of water molecules to enable a simultaneous reaction with CO$_2$ and proton donation to hydroxide.  It is unclear why the simple process of reacting CO$_2$ with zinc-bound hydroxide followed by water exchange would require such complexity.

$k_cat$ for CA II has a strong hydrogen kinetic isotope effect (KIE).\cite{silvermanProtonTransferCatalytic1983} This seems to conflict with the established mechanism of a direct nucleophillic attack of hydroxide onto CO$_2$, which should have no such KIE.\cite{pockerStoppedflowStudiesCarbon1977} The usual explanation of this apparent contradiction is that the rate limiting step of the reaction is unrelated to the bicarbonate formation reaction and must instead be associated with the deprotonation of the water molecule bound to the zinc ion to form a hydroxide. However, high-quality DFT studies show that the transition state energy barriers for the three water proton hopping mechanism are very low, such that this reaction should occur very quickly.\cite{cuiProtonWireConcerted2003}  This is also observed in the simulations here, where this reaction is  observed rapidly in equilibrium simulations. %Additionally, if this were the rate limiting step the reaction should continue to accelerate as pH increases, however, it remains flat above a pH of ~8. \cite{silvermanProtonTransferCatalytic1983}

In contrast, the reaction mechanism observed here clearly has a significant direct hydrogen kinetic isotope effect due to the coupled proton transfer. No KIE is observed for $k_\text{cat}/K_m$ experimentally but this can be attributed to the reaction becoming rate limited by the diffusion of CO$_2$ into the active site.\cite{silvermanProtonTransferCatalytic1983}

These simulations also potentially explain an additional puzzle in the literature. NMR studies show that the zinc is bound to a hydroxide ion even down to pH 5.\cite{liptonZincSolidStateNMR2004} This result is inconsistent with the traditional reaction pathway, as the hydration of CO$_2$ is halted below pH 7, which is attributed to the zinc bound water having a pKa of 7. However, another possibility becomes apparent from our simulations, which is that  below  pH 7 the additional proton is shared between the water bound hydroxide and the Glu106 residue.  When the proton is on the Glu106 residues the hydroxide bound to the zinc has a hydrogen bond donated to it from Thr199. This may prevent the OH$^-$ from being rigidly oriented towards the Thr199 and inhibit it from accepting a proton from the neighbouring water molecule and explain the low rate of the reaction below pH 7, even if hydroxide is still present. 

\subsection{Mutational studies}
Mutational studies show that Glu106 and Thr199 residues are critical to the reaction rate \cite{xueStructuralAnalysisZinc1993}. This is consistent with our simulations, where these residues act as proton acceptors during the reaction and are key to stabilizing the orientation of the hydroxide, allowing it to accept a proton from the neighboring water easily. The ability of these residues to accept protons from the zinc-bound water has been previously proposed on the basis of theoretical studies.\cite{bottoniNewModelTheoretical2004} However, it is unclear why Glu106 should play such a critical role in the established mechanism. 

A second key piece of mutational evidence is that $k_\text{cat}$ for CA I enzyme is an order of magnitude slower than CA II despite the fact that its active site is almost identical to that of of CA II, particularly around the zinc ion.\cite{silvermanProtonTransferCatalytic1983} However, there is one most significant difference between these two classes of enzymes, which is the Thr200 group is replaced with a histidine. This is the residue that stabilizes the water molecule (W2) that Orb predicts will react with the CO$_2$ in CA II. As a result, this water molecule must be missing from CA I enzymes, meaning that it must use a different reaction mechanism. This may explain this significant difference in reaction speed. 

Additionally, the most active variant of the CA XIII enzyme is the V200T mutation, resulting in a  67\% increase in catalytic efficiency. \cite{desimoneExplorationResiduesModulating2019} This again highlights the crucial role of the Thr200 residue. It is not clear why this residue should play such a critical role in the traditional reaction pathway. 

Additionally, the mutation of Asn67 and Asn62 to the hydrophobic Leu also reduces $k_\text{cat}$ substantially.\cite{fisherSpeedingProtonTransfer2007a} This is consistent with the reaction mechanism observed here as these residues could stabilize the transition state by interacting with the bicarbonate, but it is unclear why they would be needed for to accelerate a simple proton transfer.  

\subsection{Hydrophobic binding pocket}
One counterintuitive aspect of the mechanism observed here is that there is a strong binding site for carbon dioxide directly adjacent to the Zn-OH$^-$. Why would this be if it is not meant to facilitate direct CO$_2$ OH$^-$ reactions? Its possible that this pocket is not easily accessible to the CO$_2$ at high pH as it is occupied by a water molecules strongly bound to the zinc bound OH$^-$. This is consistent with our observations in Orb that the CO$_2$ reliably reacts with a water molecule in the active site before it can migrate to the hydrophobic pocket. 

We only observed spontaneous migration of CO$_2$ into this pocket when the Glu106 residue is protonated, meaning the zinc bound OH$^-$ is not rigidly oriented and can destabilise the surrounding water molecules, which are now forced to accept a hydrogen bond from the OH$^-$ group. As a result, they are oriented with their dipole in an unstable direction.  The Glu106 is more likely to be protonated at low pH, where the reverse reaction occurs (bicarbonate to CO$_2$). This may suggest that the hydrophobic pocket is important for removing the CO$_2$ from bicarbonate but not vice versa. This is consistent with the simulations above where the bicarbonate bound to the zinc easily reverts to CO$_2$. 

\subsection{Bicarbonate crystallization}
Intriguingly no one has ever been able to stabilize a native CA II enzyme with bicarbonate bound to the zinc,\cite{xueCrystallographicAnalysisThr2001993}, whereas it is possible to stabilize it with CO$_2$ in this favourable binding pocket.\cite{domsicEntrapmentCarbonDioxide2008,kimTrackingSolventProtein2016}.  There is one study that was able to crystallize a CA enzyme with bicarbonate attached to the zinc.\cite{xueCrystallographicAnalysisThr2001993} It did so by mutating the threonine group (Thr200) that stabilized the water molecule that the CO$_2$ reacts with in our simulations. It is replaced with a large histidine group, entirely displacing the water molecule. This is the same mutation that most clearly distinguishes CA I enzymes from CA II. This mutation would prevent the mechanism observed here from occurring, which is consistent with its much slower reaction rate and explains why it is required to stabilise the bicarbonate on the zinc such that it can be observed cryptographically.

\section{Conclusions and future work}
In conclusion, we have simulated the CA II enzyme using Orb, a recently developed universal neural network potential. Its ability to reproduce key known features of the reaction mechanism of this critically important enzyme, while also suggesting a new plausible reaction mechanism is a remarkable demonstration of the capability of this approach to build a more complete understanding of complex chemical reaction dynamics. 

The automatic reaction detection used here should enable this approach to be scaled to study a wide range of important systems where molecular-scale processes remain poorly understood. 

However, several important steps remain to fully validate and extend our findings: Firstly, fine-tuning the Orb potential to a higher level of DFT theory specifically for the CA II active site could improve accuracy further and hopefully enable accurate reaction rate calculations. Including quantum nuclear effects will also be important for a more quantitative description. 

Secondly, constraining the CO$_2$ to be within 5 \AA\ of the zinc is a crude method. To build a more complete mechanistic picture and account for the role of CO$_2$ diffusivity properly longer simulations with more solvating waters and more sophisticated enhanced sampling techniques such as metadynamics or umbrella sampling should be used. These approaches will enable calculation of precise free energy profiles along the reaction coordinate. 

Thirdly, simulating additional CA variants with known differences in catalytic rates could help validate our mechanism by correlating structural features with experimental kinetics. Simulations of CA mutants with altered activity could provide predictive power for enzyme design and medical diagnosis and treatment. A particularly important case is the replacement of Thr200 with histidine that is the main distinction between CA I and CA II enzymes. 

This approach could also be useful for filtering candidate synthetic enzymes designed using generative models such as RFDiffusion.\cite{watsonNovoDesignProtein2023} An analogous role was performed with ChemNet\cite{anishchenkoModelingProteinsmallMolecule2024} in a recent article outlining the computational design of novel enzymes.\cite{laukoComputationalDesignSerine2025} Simulations of the enzyme with known small molecule inhibitors could also provide important insights into drug discovery strategies.

Perhaps most excitingly, our findings suggest new directions for designing artificial CO$_2$ capture systems. Specifically, it may be possible to evaluate potential synthetic catalysts that could exploit the water-mediated CO$_2$ conversion mechanism we observed. 

\section{Acknowledgment}
We would like to acknowledge the entire team at Orbital Materials for support, assistance and computational resources. Mark Neumann, Matthew Kapelewski, Corin Wagen and  Alexander Mathiasen are also acknowledged for helpful detailed feedback. Thanks also to Alexander Mathiasen for the development of the logmd package.  

\section{Data availability}
All code presented here is available at: \href{https://github.com/timduignan/CAII-Orb-d3-v2}{github.com/timduignan/CAII-Orb-d3-v2}   \newline
Trajectory of the proton hopping relay to His64 is available at: \newline
\href{https://rcsb.ai/bf7e353ddf}{rcsb.ai/bf7e353ddf}  (Timestep: 10 fs) \newline
Trajectory of the bicarbonate formation is viewable at: \newline
\href{https://rcsb.ai/e138d28fe8}{rcsb.ai/e138d28fe8}
  (Timestep: 10 fs) \newline
Trajectory of the proton hopping to the Glu 106 after the bicarbonate formation is available at: \newline
\href{https://rcsb.ai/ddf302794a}{rcsb.ai/ddf302794a}
 (Timestep: 10 fs) \newline
Trajectory of the CO$_2$ migrating into the hydrophobic pocked is available at: \newline
\href{https://rcsb.ai/aa0b3bf874}{rcsb.ai/aa0b3bf874}
 (Timestep: 5 ps) \newline
Trajectory of the bicarbonate bound to the is available at: \newline
\href{https://rcsb.ai/96ae247a31}{rcsb.ai/96ae247a31}  (Timestep: 5 ps) \newline
These were created using the logmd package.\cite{mathiasenHttpsPypiorgProject2025}

\newpage
\printbibliography

\end{document}